# Anisotropic magnetic nanoparticles for biomedicine: bridging frequency separated AC-field controlled domains of actuation

*David Serantes[1,2], Roy Chantrell[2], Helena Gavilán,[3] María del Puerto Morales[3], Oksana Chubykalo-Fesenko[3], Daniel Baldomir[1] and Akira Satoh[4]*

[1]Applied Physics Department and Instituto de Investigacións Tecnolóxicas, Universidade de Santiago de Compostela, 15782, Spain

[2]Department of Physics, University of York, Heslington, York YO10 5DD, United Kingdom

[3]Instituto de Ciencia de Materiales de Madrid, CSIC, Cantoblanco, ES-28049 Madrid, Spain

[4]Faculty of System Science and Technology, Akita Prefecture University, Yuri-honjo 015-0055, Japan

**ABSTRACT**

**Magnetic nanoparticles constitute potential nanomedicine tools based on the possibility to obtain different responses (translation, rotation, heating, etc.) triggered by safe remote stimulus. However, such richness can be detrimental if the different performances are not accurately differentiated – and controlled. An example of this is the reorientation of magnetic nanoparticles under the influence of AC fields, which can be exploited for either magneto-mechanical actuation (MMA) at low frequencies (tens of Hz); or heat release at large ones (0.1 – 1 MHz range). While it is clear that Brownian rotation is responsible for MMA, its heating role in the high-frequency regime is not clear. In this work we aim to shed light on this issue, which needs to be well understood for applications in magnetic**

**fluid hyperthermia (MFH) or heat triggered drug release. Using a Brownian dynamics (BD) simulation technique, we have theoretically investigated the contribution of Brownian reversal in disk-shape particles (to enhance the viscous interaction with the environment) over a wide range of frequencies. Our results predict essentially negligible hysteresis losses both in the high- and low-frequency domains, with completely different implications: highly efficient MMA, but negligible MFH performance. Importantly, complementary micromagnetic simulations indicate that the large magnetic torque assumption of the BD simulations is supported by hexagonal-shape disks, up to field amplitudes of the order of 100 Oe. Larger fields would lead to Néel reversal which, noteworthy, predicts significant heating performance. The possibility of switching between the MMA and MFH response by changing the amplitude of the AC field, together with their distinct optimal conditions (large magnetic torque for MMF; large heating for MFH), points to such hexagonal nanodisks as promising nanomedicine agents with double mechanical and heating functionalities.**

Magnetic nanoparticles are promising biomedical agents due to their versatile response under different external stimulus, with a number of potential applications that range from drug delivery, imaging, hyperthermia cancer treatment or purely magneto-mechanically actuated responses (e.g. for breaking blood clots).[1] Such different applications are treated as separate research areas for the related communities based on their differentiated actuation requisites.

Nevertheless, in recent years there has been a growing attention to the development of nanoentities with different functionalities, so that various actuations could be achieved within the biological tissue by using the very same system. The main example is theranostics, where the same agent can be used for both detection and treatment.[2] Such multiple fields of actuation by the same entity require an accurate control of the triggered responses. This is straightforwardly achieved if the required actions are promoted by different types of fields, as e.g. drug delivery and release (large homogeneous field gradients *vs.* AC-field triggered thermal release of the drug cargo),[3] but has the disadvantage that the engineering apparatus is far more complex. It is clear that being able to have differentiated actuations while using the same apparatus would have great advantages. In this respect, magnetic fluid hyperthermia (MFH) and magneto-mechanical actuation (MMA) seem ideal candidates, since the working principle is the same (excitation by an external AC magnetic field) the differentiation arising from the different time domains of the phenomena.[4] In principle, simply by changing the frequency of the AC field it is possible to disentangle both responses, obtaining either magnetic actuation (Brownian regime, at low frequencies) or heating (Néel reversal regime, at high frequency). At this point it is important to emphasize that although some works still try to distinguish the contribution of Brownian reversal to hyperthermia (in the context of fitting to the Linear Response Theory, whose limited applicability to MFH has been well documented[5]), it has been repeatedly shown that Brownian rotation plays a negligible role for heat dissipation under biological conditions.[6,7,8] Also, neither has the mechanical response of the Brownian rotation in the high-frequency MFH regime proved to cause significant cell damage,[9] in agreement with the observation that cellular internalisation can disable Brownian relaxation.[8]

In this context, we have recently shown that even if the Brownian contribution can be discarded as heating mechanism, it may still play a role in tuning the heat dissipation by Néel reversal in chains of spherical particles[10] and nanorods[11]. This occurs because there is a progressive reorientation of the elongated structures during the AC treatment (parallel to the field in the case of the chains; orthogonal in the case of the rods) that allows for enhanced performance of Néel-type heating.[12] Therefore the question is; is such reorientation really accompanied by negligible contributions to the overall heating? Understanding the heat contribution of anisotropic nanomagnets over a wide range of frequencies (from the MMA regime to the purely MFH one) itself constitutes a crucial aspect to solve towards the possible use of anisotropic nanomagnets for combined mechanical actuation and heating, either sequentially or simultaneously.

In order to shed some light on this question, we have theoretically investigated what would be the heating properties of anisotropic nanomagnets in a viscous environment, if the Néel contribution is quenched, for example in large (> 20 – 50 nm) nanoparticles with high anisotropy. By using a Brownian dynamics technique[13] we have simulated the AC field response of magnetic disks of large (> 2 - 4) aspect ratio, able to provide large mechanical action. The interest in the disk shape is the similarity to the blood cells freely travelling within the vessels, hence constituting a very interesting morphology for the colloidal properties for intravenous injection. The particle magnetization is assumed to lie in the plane of the disks, and to be strongly coupled to the lattice so that particle reorientation is directly link to magnetization rotation (see Computational Details section). This negligible thermal diffusion is due to the large particle size, which also prevents superparamagnetic fluctuations hence ensuring the delivery of a large magnetic torque.[14] The particle properties are those of magnetite, based on the suitability of iron oxides for biomedicine in comparison with other nanoparticles.[15] Particularly, as reference we have used disks of same dimensions and composition as those recently reported by H. Gavilán *et al.* [16] obtained by dry reduction of an antiferromagnetic precursor, i.e. uniform magnetite particles of hexagonal shape with average dimensions 22 nm thickness and 140 nm diameter with a standard deviation of 10%. The experimental hexagonal shape is not the same as assumed in the BD simulations because of the difficulty of taking into account irregular shapes in the BD treatment; however we note that the experimental samples are coated with a silica shell that softens the edges of the hexagons (regarding the viscous interaction with the environment) and hence the disk shape constitutes a rather good approximation to the experimental one. Sketches illustrating the shape of the disks assumed in the BD simulations are shown in Figure 1, together with the TEM image of the experimental samples from reference [16], both the as-prepared (A), and silica-coated (B) with smooth shapes. Importantly, those particles have a large in-plane magnetic moment, suitable for the BD description, as will be discussed later.

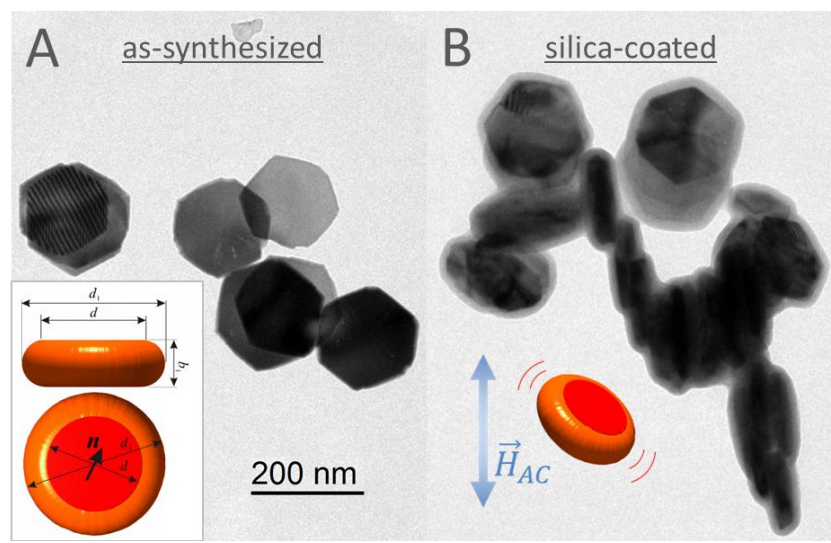

**Figure 1.** A: Drawing sketches of the simulated particles (in red), together with some SEM examples of the experimental samples taken as reference, either as-prepared (A) or silica-coated (B).[16] The scheme on the right aims to illustrate that the effect of the AC field is to promote physical reorientation.

It is known that not only large static fields, but also small alternating ones can cause aggregation,[17] consequently enhancing the role of interactions. Nevertheless, since in this work we are interested in the pure Brownian effects of the anisotropic particles with the environment, interactions with other particles will not be considered here (neither magnetic nor mechanical effects). Thus, the investigation procedure is to simulate the magnetization vs. field hysteresis loops, *M(H(t))*, of a random distribution of disks under the influence of an alternating AC field. Parameters to investigate are the role of the media viscosity, $\eta$; the amplitude ($H_{max}$) and frequency (*f*) of the AC field, given as $H_{AC}(t)=H_{max} sin(2\pi f \cdot t)$; and the size and aspect ratio of the particle.

Some examples of the time evolution of the magnetization with the varying field are shown in Figure 2A for different frequencies, for a viscosity $\eta = 2.0\times10^{-3}$ Pa·s (at room temperature) corresponding to that within HeLa cells (commonly used in cancer investigations). At this point it is worth emphasizing, however, that when dealing with nanosized dimensions the viscosity is not just defined by the solution properties, but also by the probe size, leading to definitions of "nano" viscosities for small hydrodynamic radius of the probes (order of tens of nanometres) and "macro" viscosity for larger sizes.[18] While this factor is not the main aim of our current work, it is clearly of crucial importance for the related results and hence we decided to reflect it in our investigation. Therefore we have considered both the "nano" and "macro" viscosities of HeLa cells, which according to T. Kalwarczyk *et al.* [18] have values of $\eta_{nano} = 2.0\times10^{-3}$ and $\eta_{macroo} = 4.4\times10^{-2}$ Pa·s, respectively. The systematic analysis of the energy losses of the studied disks for those viscosity values are shown in Figure 2B, together with that of water, included as a reference.

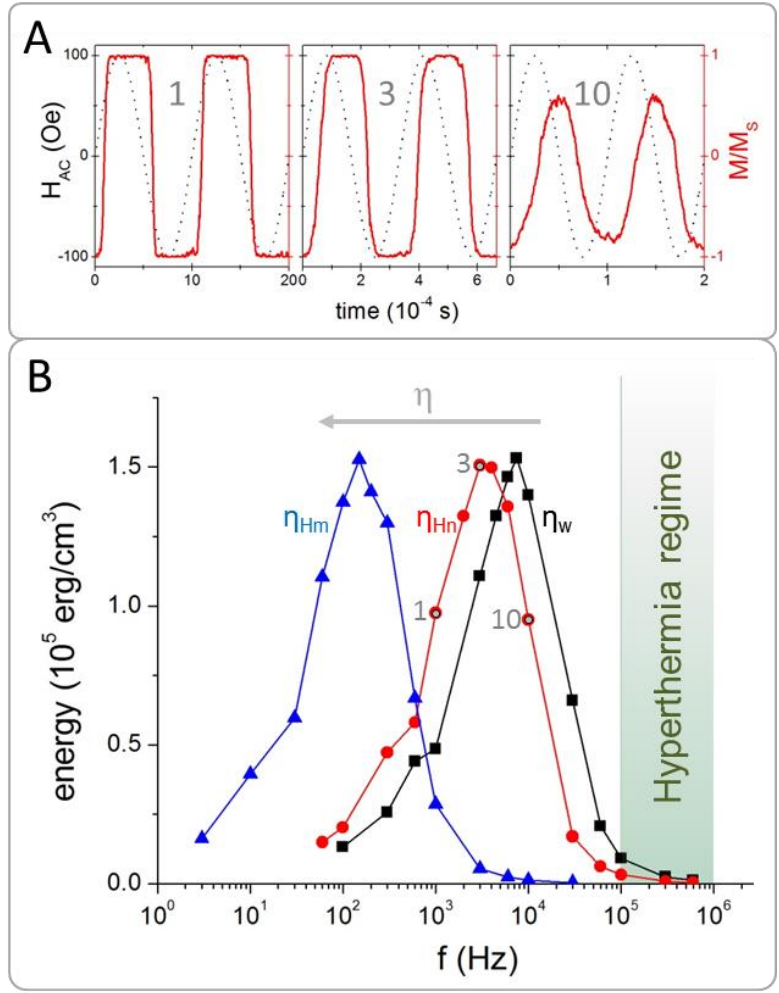

**Figure 2.** A: time evolution of the magnetization (red solid line) under an AC field (black dotted line) of $H_{max}$ = 100 Oe, at room temperature and for different frequencies ($f$ = 1, 3, and 10 kHz, respectively), for HeLa "nano" viscosity, $\eta_{Hn}$ = 2.0×10$^{-3}$ Pa·s; B: energy losses as a function frequency, for 3 different viscosities at RT: water (black squares), $\eta_w$ = 0.89×10$^{-3}$ Pa·s; and HeLa cells, both for the nano (red circles) and macro (blue triangles); $\eta_{Hm}$ = 4.4×10$^{-2}$ Pa·s. The direction of increasing viscosity is indicated by the horizontal grey arrow. The points 1, 3, 10 correspond to the data in B (in kHz), and the shaded area illustrates the frequency regime usually considered in MFH experiments.

The results displayed in Figure 2B show a common trend for the 3 viscosities investigated: a Gaussian-like shape which peak shifts towards smaller frequencies for higher viscosities. Notably, the hysteresis losses (HL) become very small at high frequencies, supporting the negligible role of Brownian contribution for MFH[6-8] and with the detrimental effect on the heating of increasing viscosity for large anisotropies.[19] On the opposite side of the peak, at low frequencies, the losses become also negligible when reaching values as low as tens of Hz, i.e. the values reported as efficient for causing membrane integrity loss by MMA using magnetic microdisks.[20] Note that for efficient MMA a large magnetic torque is required, and such values are achieved for a strong coupling between magnetization and lattice, i.e. no phase lag between magnetization and field. Then, depending on the viscosity of the environment and field conditions, that torque can lead to a large reorientation amplitude for mechanical action at low frequencies (magnetization switches completely with the field, i.e. full particle reorientation is reached), or just to a small oscillation amplitude at large frequencies (magnetization unable to switch, small changes). Finally, it is important to note that the maximum HL values reached are of the order of the particle anisotropy, emphasizing that both Néel and Brown relaxation processes could potentially lead to similar heating performance (note that Néel hyperthermia performance is directly proportional to the particles' anisotropy, as illustrated by the *magnetic hyperthermia trilemma*[21]). Regarding attainable energy losses, the role played by the viscosity seems to be to shift the peak to lower frequencies, without changing the maximum value.

The results in Figure 2 were obtained for a particular value of the AC field amplitude ($H_{max}$ = 100 Oe), but it is reasonable to expect that increasing its value could increase the Brownian heating at larger frequencies. In order to check whether such an increase could lead to the attainment of a more significant fraction of the maximum achievable value, we have investigated the role of the field amplitude. Values of the energy losses for various values of $H_{max}$ (of the same order as those used in the MFH experiments) are plotted in Figure 3, for the case of "nano" viscosity of HeLa cells.

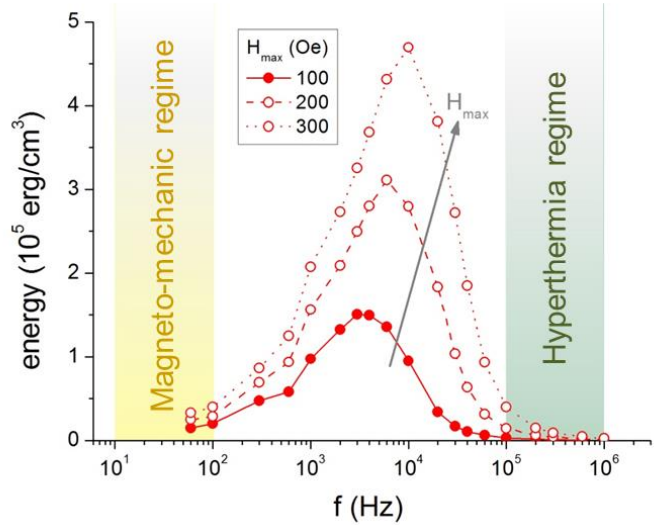

**Figure 3.** Energy losses *vs.* frequency, for different $H_{max}$ values. The reference viscosity is that of HeLa cells (nano), i.e. $\eta_{nano} = 2.0 \times 10^{-3}$ Pa·s. The shaded areas illustrated the usual regimes for MMA (low-f) and MFH (high-f), respectively.

The results shown in Figure 3 clearly indicate an overall growth of the energy losses with larger fields for all frequencies, but not significantly in the hyperthermia regime. However, while such values remain quite low in comparison with the maximum possible values, it is important to compare them with the energy that could be obtained by Néel reversal. In this regard, for the comparison it is important to consider not only the relative performance of both Brownian and Néel contributions, but also the field that would trigger the Néel switching. Note the different nature of both mechanisms regarding heating: while the Brownian contribution continuously grows with $H_{max}$, Néel reversal has a significant threshold related to the anisotropy field and larger AC field amplitudes do not provide additional heating.[10,12] In the following the Néel heating properties will be studied.

The large aspect ratio of the disks, which suggests a strong role of the shape anisotropy on the reversal of the magnetisation, could however be detrimental if allowing for non-coherent reversal modes. This does not seem to be the case with the hexagonal particles taken as a reference, which show a rather uniform magnetization state.[16] In order to confirm such coherent-like behaviour and evaluate the corresponding hysteresis losses due to Néel reversal, we have used micromagnetic simulations (using the OOMMF software package[22]). Since a random dispersion is expected to exist in the crowded biological matrices, we considered different directions of application of the field over the particle and studied how the energy losses depend on the angle $\theta$, as illustrated in Figure 4D as reference system. Note that while for the BD modeling we used the rounded-edges shape described in Figure 1, which would resemble the magnetic particle with the silica coating (TEM image in Figure 1B), for the simulation of the reversal process only the magnetic part is relevant. Therefore it is necessary to take into account the exact hexagonal shape of the nanomagnets (TEM image in Figure 1A), as will be shown below. The simulation values used correspond to those of the magnetite disks of Ref. [16] (see Computational Details section). The micromagnetic results are summarized in Figure 4.

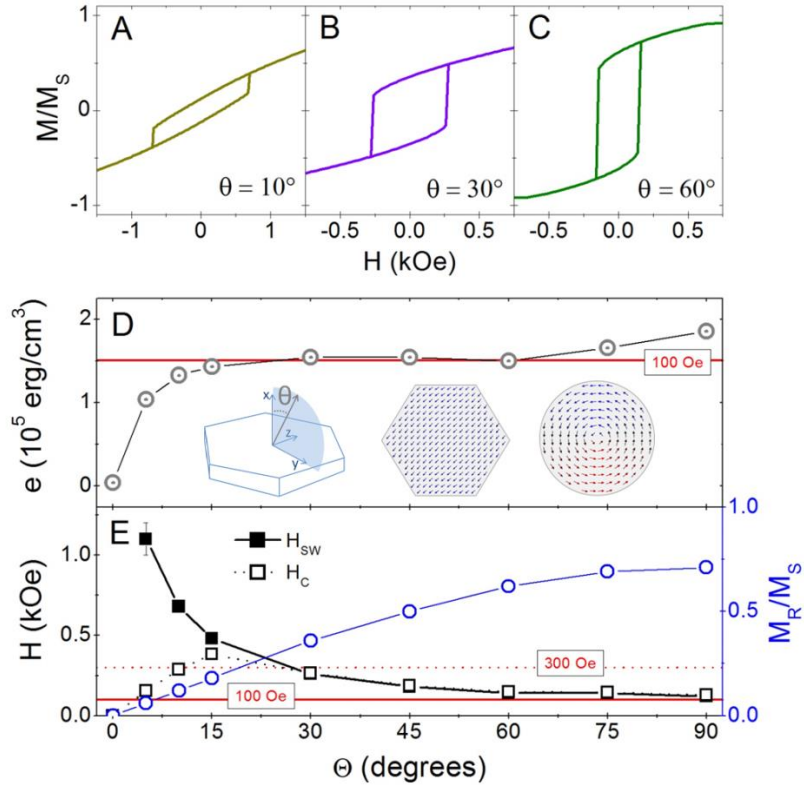

**Figure 4.** A, B, C: examples of *M(H)* hysteresis loops at different angles between field and hexagonal disk (angle orientation is illustrated in D). D: energy losses (major loops) vs. angle; the horizontal red line corresponds to the maximum energy losses by Brownian reorientation at 100 Oe, as a reference. The drawing indicates the reference of direction of field application, and the snapshots illustrate magnetisation configuration at remanence, for hexagonal and disk shape, respectively. E: Coercive ($H_C$) and switching ($H_{SW}$) fields, and remanence ($M_R$) as a function of angle. Horizontal red lines correspond to 100 Oe and 300 Oe, respectively.

The *M(H)* hysteresis loops shown in Figure 4 A-C illustrate the change in magnetic behaviour depending on the angle between applied field and particle direction, θ. Importantly, the magnetic moment of the particle remains essentially unchanged throughout the entire loop (less than 1% decrease for all values of θ and field values), in agreement with the experimental observations on the uniformity of the magnetization.[16] Such quasi-coherent behaviour is indicative of a large torque for MMA, which will promote Brownian rotation of the particles unless the field amplitude is large enough to lead to field-driven (Néel) switching over the internal energy barrier. In this regard, the overall shape of the loops is quite similar for all values of θ, with an abrupt jump that corresponds to the irreversible switching of the magnetisation ($H_{SW}$) and suggests uniaxial symmetry for the anisotropy barrier. The variation of both $H_{SW}$ and remanence ($M_R$) with θ suggests different hysteresis losses, as systematically analysed in Figure 4D. Interestingly, the energy loss follows a rapid increase from low angles and then becomes essentially constant (angle-independent), which could be interesting for diminishing the dispersion in local heating.[23] This behaviour reinforces the interest of such magnetic nanostructures for MFH; however, this is not the main objective of the current work so for the moment we focus our attention on the competition with the Brownian-origin heating performance. In this regard, it is worth noting that the red horizontal line in Figure 4D stands for the peak of the hysteresis loss *vs.* frequency due to Brownian relaxation (Figure 3), pointing out the similarity in the maximum achievable heating performances from both Néel and Brownian relaxation. While the Néel losses cannot be further increased (being defined by the particle anisotropy), the Brownian losses in principle can be continuously increased with larger field amplitudes, as illustrated in Figure 3. Yet, it is important to emphasize that in the MFH regime, the Brownian energy losses only reach a small fraction of the peak value, thus anticipating a negligible contribution in relation to the Néel losses, as expected.[6-8] The snapshots of the hexagonal and disk shapes correspond to typical magnetization configurations at remanence for both shapes, emphasizing the importance of accurate shape description for the magnetization reorientation description: circular shape allows for vortex-like configuration, with very small magnetic moment (and hence negligible torque), very different than experimentally observed with the hexagonal samples.

In order to further investigate the heating performance of either Brownian or Néel reversal, in Figure 4E the evolution of $H_{SW}$ *vs.* θ is shown, since the switching field represents, to a first approximation, the minimum threshold for Néel-type magnetization reversal. Interestingly, plotting also the coercive field, $H_C$, we observe a correlation of the angle-independent hysteresis loss range with the overlapping of $H_{SW}$ and $H_C$, for values θ > 15º. Regarding the $H_{SW}(θ)$ evolution, this shows a monotonic decrease starting from θ =5º (no switching for the anhysteretic θ = 0º case), approaching asymptotically the 100 Oe value (solid horizontal red line) with increasing angle. This suggests that Néel reversal could be negligible for fields of the order of or smaller than 100 Oe, resulting in domination of Brownian rotation. Note that this is the simplest possible interpretation, since thermal effects could help in promoting the reversal at smaller fields: an effect which will become increasingly important for small particle size.[24] In our current approximation, these results suggest that at larger fields (300 Oe is also depicted as an horizontal dotted red line) Néel reversal could dominate over Brownian rotation, depending on the relative orientation with respect to the applied field. Since the choice of either path would be decided by energy costs, this demonstrates the importance of checking the energy dissipated by particles at different orientations either by Néel or Brown reversal, for different $H_{max}$ values.

Comparing Néel and Brownian contributions to heating may be complicated, since both processes are expected to coexist. Furthermore, the progressive reorientation *via* Brownian rotation would dynamically change the Néel contribution, leading to the possibility that eventually all particles could reorientate up to a point at which Néel reversal is more favourable. While these arguments are sound, it is also important to take into account that the time scales of Brownian reorientation and Néel reversal are very different (orders of magnitude), particularly with the current particle sizes.[25] Furthermore, we have previously shown that while the Brownian relaxation time would be at most of the order of milliseconds, its effects may be macroscopically observed in MFH experiments –indirectly, through the change in the Néel heating- in the timescale of minutes due to rotation of the easy axes.[11] Hence, for a system initially at random, it makes sense to evaluate both contributions as separate ones. Therefore and for the sake of simplicity, we have decided to consider a random distribution of particles and assume that the fraction of those dissipating by Néel reversal is that for which $H_{max} > H_{SW}$. This is illustrated in Figure 5, where the MFH frequency regime from Figure 3 is reproduced and augmented, to be taken as a reference for the hyperthermia performance.

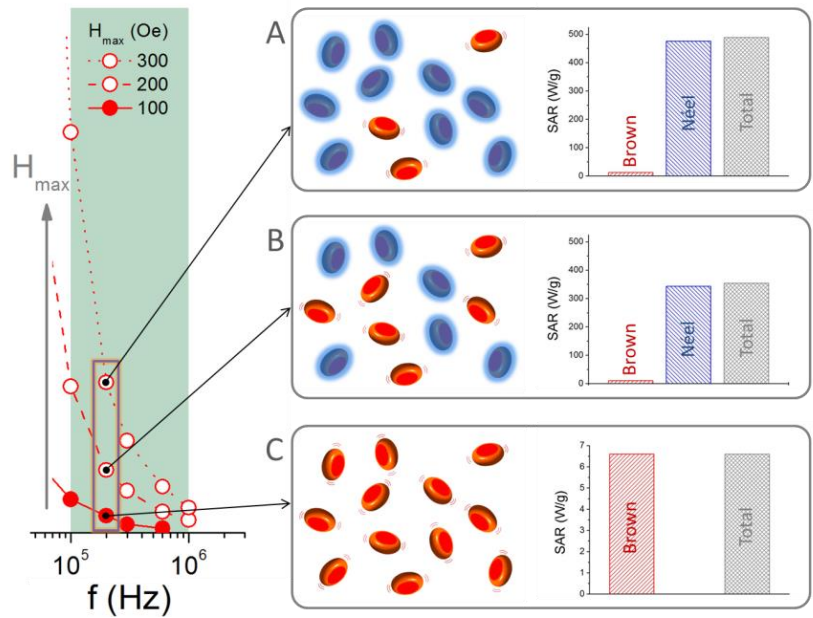

**Figure 5.** Estimation of the *Specific Absorption Rate* (SAR), for the $f$ =200 kHz case of Figure 3, for 3 different field amplitudes (A = 300 Oe; B = 200 Oe, C = 100 Oe); note the scale difference between A and B (the same) with respect to C (much smaller). The drawings illustrate heating *via* Brownian or Néel reversal (red and blue disks, respectively).

Figure 5 illustrates the relative importance of Brownian or Néel reversal contributions to the heating, under the simple approximation approach described above: if the AC field is not able to switch the magnetization *via* Néel reversal, then Brownian reorientation is the heating source. Since the switching field is defined by the relative angle between AC field and particle, then in this simple picture it is possible to estimate the fraction of particles that prefer Néel reorientation ($n_{\text{Néel}}$) by correlating $H_{\text{max}}$ and $H_{\text{SW}}$ (as shown in Figure 4E), so that $n_{\text{Néel}} = \int_{\alpha}^{\pi/2} \cos\theta \sin\theta \, d\theta$, where α is the angle at which $H_{\text{max}} = H_{\text{SW}}$. From Figure 4E it can be estimated α ≈ 41° (200 Oe) and 27° (300 Oe), so that the fraction of particles that may reverse by Néel increases from no contribution (at $H_{\text{max}}$ = 100 Oe), to 0.57 and 0.79 for $H_{\text{max}}$ = 200 and 300 Oe. Then, taking into account the heating performance of each type of particle, it is possible to estimate the total performance of each type of reversal for the entire system. This calculation is performed in terms of the *Specific Absorption Rate* (SAR), SAR = HL*frequency, the standard parameter for MFH. For the Brownian contribution, $SAR_B$, this value depends on $H_{\text{max}}$ (as shown in Figure 3), estimated in average for a single particle $SAR_B$ = 7, 25 and 60 W/g; for $H_{\text{max}}$ = 100, 200 and 300 Oe, respectively. The Néel part, $SAR_N$, is taken as the average constant value shown in Figure 4D, estimated as $SAR_N$ = 603 W/g (and, as explained above, it does not depend on $H_{\text{max}}$). Then, taking into account also the relative fraction of each type of reversal (illustrated in the sketches as red or blue colours for Brownian and Néel reorientation, respectively), it is possible to estimate their contributions to heating as a function of $H_{\text{max}}$. The results are shown as histograms in Figure 5, where it is observed a large difference in the SAR value depending on the field amplitude. For $H_{\text{max}}$ = 100 Oe, when only Brownian rotation contributes to heating, the reached SAR values are very small (a few watts per gram); on the contrary, when Néel reversal becomes the main contribution to heating (for larger $H_{max}$ values) a much larger SAR is reached. These results clearly demonstrate the negligible role of Brownian reversal for heat dissipation, in agreement with Refs. [6-8], except for the very specific (inefficient) low-$H_{max}$ conditions. Furthermore, it is worth emphasizing that the SAR values predicted (of the order of 500 W/g for $H_{max}$ = 300 Oe and f = 200 kHz) are quite significant, pointing to a potential use of the investigated disks as heat generators for MFH. In addition, the clear separation of their performance into specific frequency regimes, together with their robust magnetic behaviour, suggests also the possibility of mechanical actuation. This suggests the intriguing possibility of alternating between purely magnetic and mechanical actuation to optimize the therapeutic results. As an additional comment, we would like to point out that taking into account the large timescale difference between Néel and Brownian processes, it may be important, when estimating the SAR value from the initial slope of the temperature increase *vs.* time curve, to consider not only the heating mechanism, but also the fraction of the system actually dissipating significant energy.

The above considerations are strongly dependent on the fact that the internal magnetization processes behave coherently, which based on the micromagnetic simulations seems to be strongly determined by the particular hexagonal shape. Therefore, for additional studies it would be interesting to perform similar analyses as a function of particle size/aspect ratio, and also for other shapes. While such systematic analysis is not the objective of the current work, it may still be worth investigating the role of size on the Brownian Dynamics results. The motivation for this is the fact that we assumed a highly symmetrical shape for the disks (Figure 1), despite the hexagonal magnetic part (as-prepared disks, TEM images of the experimental samples, Figure 1A), under the assumption that the silica coating would smooth the sharp edges of the hexagons (TEM images of the experimental samples, Figure 1B). But, if a silica shell is present, then the absolute size to compare the Brownian and Néel contributions should not be the same. In order to have an initial view on the role of the absolute size on the BD results, in Figure 6 the hysteresis loss *vs. f* data is shown (as in Figures 2 and 3), for the "nano" viscosity and $H_{max}$ = 100 Oe, for 2 different sizes ranging from 100 to 200 nm in diameter, smaller and larger than the used one, 140 nm. The effect of the disc size is clearly negligible, which gives additional support to our interpretation.

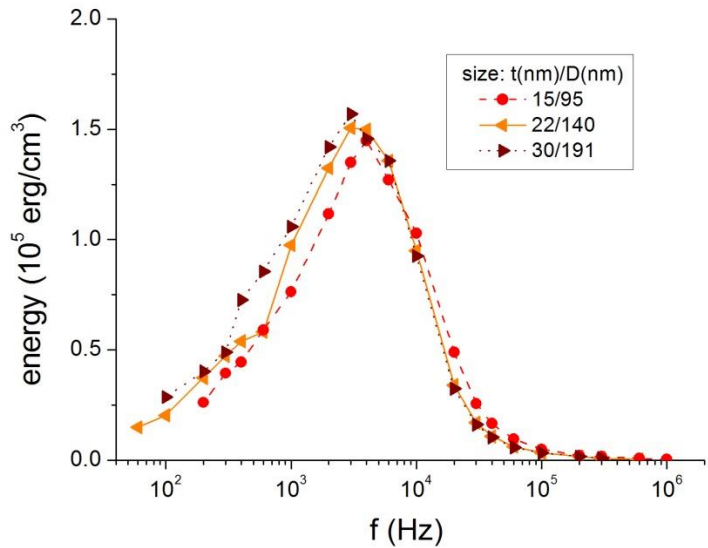

**Figure 6.** Energy losses *vs.* frequency, for the same field and viscosity conditions as in Figure 3 ($H_{max}$ = 100 Oe case), for different absolute sizes of the particles but same aspect ratio.

Our combined Brownian dynamics/micromagnetic study indicates a negligible role of Brownian rotation for heat dissipation in the magnetic hyperthermia range of frequencies, in agreement with some experiments.[6-8] Furthermore, the fact that we have focused on highly anisotropic particles (to maximize the viscous interaction with the environment and hence maximize the losses), suggests that for most experimental samples –usually with spherical particles-, the Brownian contribution to heating is even smaller, at least for coherent-like magnetization reversal process. At this point we consider, on the basis of our claculations, the *effective* relaxation time approach followed by several insightful works[26] for the description of the particles' behaviour, under the assumption that both Brown and Néel reorientation would coexist. Such approach is usually applied to particles in the 10 – 20 nm size, where both relaxation times can be of the same order of magnitude. However, a quick estimation of the Néel relaxation time for the large sizes considered in the current work yields values of the order of years, hence ensuring a complete inhibition of Néel relaxation and giving additional support to our approach on the separation of the heating contributions. Nevertheless, it is still necessary to keep in mind the need for further refinements, since timescale of minutes for easy-axes reorientation should certainly be taken into account. It is also important to mention here that while the Brownian process has taken into account proper dynamics and thermal effects, the micromagnetic simulations did not consider thermal fluctuations, leading to the possibility that in the real system the Néel reversal could take place at smaller fields (thermally-assisted switching). Further studies should take into account such considerations, which were however secondary to the current work, dealing with the study of the heating performance of Brownian rotation in a wide range of frequencies, and comparison with the purely magnetic dissipative contribution.

To summarize, our Brownian Dynamics simulation results indicate a clear separation of two negligible hysteresis-losses regimes for disk-shape particles, corresponding to the high-frequency *magnetic fluid hyperthermia* regime (0.1 – 1 MHz range), and to the low-frequency (tens of Hz) *magneto-mechanical actuation* regime, respectively. The low-frequency anhysteretic regime points to suitable use of those particles as medical nanorobots,[27] for a broad range of therapies based on pure nano-mechanical actuation.[28] Interestingly, the complementary micromagnetic simulations predict very significant heating properties of the investigated hexagonal disks in the *magnetic fluid hyperthermia regime* (which, in addition to large large heating power, also display an essentially angle-independent performance strongly promising for homogenous local heat generation). The fact that the hyperthermia and the magneto-mechanic actuation regimes belong to well differentiated domains, and the easy alternation between both modalities just by changing the AC field, strongly suggest the use of the magnetic nanodisks for multimodal therapeutic purposes.

**Computational details.**

**Brownian dynamics simulations.** We used the same model as in Ref. [13], i.e. disk-like particles as schematically drawn in Figure 1 which, in addition to the usual translational and rotational Brownian motion, it is considered the spin Brownian motion about the particle axis. For the characteristic parameters we used the experimental value of $M_S$ = 415 emu/cm$^3$ (300 K), and for the anisotropy we used the bulk value $K$ = -1.1×10$^4$ erg/cm$^3$. In the following we briefly summarize the basic equations for the present Brownian dynamics simulations; since there are no interactions and the AC field is homogeneous in space, only the rotational motion needs to be taken into account.

The rotational motion about a line normal to the particle axis and the spin rotational motion about the particle axis are expressed, respectively, as

$$\mathbf{e}(t+\Delta t) = \mathbf{e}(t) + \frac{1}{kT} D_\perp^R \mathbf{T}_\perp^P(t) \times \mathbf{e}\Delta t + \Delta\phi_{\perp 1}^B \mathbf{e}_{\perp 1} + \Delta\phi_{\perp 2}^B \mathbf{e}_{\perp 2} \qquad (1)$$

$$\mathbf{n}(t + \Delta t) = \mathbf{n}(t) + \frac{1}{kT} D_{\parallel}^{R} \mathbf{T}_{\parallel}^{P}(t) \times \mathbf{n}\Delta t + \Delta\phi_{\parallel}^{B} \mathbf{e} \times \mathbf{n} \qquad (2)$$

Expressions (1) and (2) show the equations providing the particle direction **e** and the magnetic moment direction **n**, respectively (note the simplification regarding the same expressions in Ref. [13], since no shear flow is considered in the present work). The components normal and parallel to the particle axis are denoted by the subscripts $\perp$ and $\parallel$, respectively; and $\Delta t$, $k$ and $T$ are the time interval, Boltzmann's constant and liquid temperature, respectively. It is noted that $D^R$ is the rotational diffusion coefficient of the disk-like particle which contains the viscosity ($\eta$), given by $D^R = \frac{kT}{\pi d_1^3 \eta} \Gamma_f$, where $d_1$ defines the absolute particle size (see Figure 1), and $\Gamma_f$ is a geometric factor that takes into account the aspect ratio of the oblate spheroid, different for the normal and parallel components.[29] The dipole-field interaction torque is given by $\mathbf{T}^P = M_S V H_{max} \sin(2\pi f \cdot t) \mathbf{n} \times \mathbf{h}$, where V is the particle volume and **h** stands for the unit vector along the field direction; if we have to take into account magnetic and steric particle-particle interactions, the torque $\mathbf{T}^P$ should include these factors. Moreover, $\mathbf{e}_{\perp 1}$ and $\mathbf{e}_{\perp 2}$ are the unit vectors normal to each other in the plane normal to the particle axis. The rotational Brownian motion is induced by the random angular displacements $\Delta\phi_{\parallel}^B$, $\Delta\phi_{\perp 1}^B$ and $\Delta\phi_{\perp 2}^B$. These quantities have the following stochastic characteristics:

$$\langle \Delta\phi_{\parallel}^B \rangle = \langle \Delta\phi_{\perp 1}^B \rangle = \langle \Delta\phi_{\perp 2}^B \rangle = 0, \quad \langle (\Delta\phi_{\parallel}^B)^2 \rangle = 2D_{\parallel}^R \Delta t,$$

$$\langle (\Delta\phi_{\perp 1}^B)^2 \rangle = \langle (\Delta\phi_{\perp 2}^B)^2 \rangle = 2D_{\perp}^R \Delta t \qquad (3)$$

in which $\langle - \rangle$ is the mean value of the quantity of interest. In the present study, the expressions for the diffusion coefficient of an oblate spheroidal particle with the corresponding aspect ratio were employed for simulations.

**Micromagnetic simulations.** We used OOMMF software package[22] to simulate the quasistatic hysteresis properties of the hexagonal disk-shape particles. The materials parameters are those of the magnetite (cubic anisotropy, $K=-0.011\times10^5$ erg/cm$^3$; exchange stiffness $A_{exch}=1.0\times10^{-6}$ erg/cm), and the saturation magnetisation $M_S = 415$ emu/cm$^3$ which was taken from Ref. [16]. The easy axes directions are (1 0 0) and (0 1 0), and the applied field was varied at small intervals by energy minimization, until the torque criteria $\boldsymbol{m} \times \boldsymbol{H} \times \boldsymbol{m} = 0.1$ is met (**m** is the reduced magnetization, **m** = **M**/M$_S$).

AUTHOR INFORMATION

**Corresponding Author**

*E-mail: david.serantes@usc.es; david.serantes@york.ac.uk

**Author Contributions**

Based on an original idea by M.P.M., the research plan of the present theoretical work was designed by D.S., R.C. and A.S. The simulations were performed by D.S., with assistance of A.S. (Brownian dynamics) and O.C.-F. (micromagnetics). The magnetic particles used as reference were synthesized by H. G. and M. P. M. Discussions and interpretations of the results were done by all authors. D.S. and R.C wrote the core of the manuscript, with input and corrections from all co-authors.

**Notes**

The authors declare no competing financial interest.

**ACKNOWLEDGMENT**

This work was partially supported by the EU project NanoMag 604448; the Spanish Ministry of Economy and Competitiveness (MAT2016-76824-C3-1-R); the Royal Society International Exchanges Scheme (IE160535); and Xunta de Galicia (GRC 2014/013; and financial support of D.S. under Plan I2C).